\documentclass{pnastwo}

	\usepackage{graphicx}
	\usepackage{amssymb,amsfonts,amsmath}
	\usepackage{subfigure}

\contributor{Submitted to Proceedings
of the National Academy of Sciences of the United States of America}
\url{http://arxiv.org}
\copyrightyear{2008}
\issuedate{Issue Date}
\volume{Volume}
\issuenumber{Issue Number}

\begin{document}

	\title{Work and information processing in a solvable model of Maxwell's demon}
	\author{Dibyendu Mandal\affil{1}{Department of Physics, University of Maryland, College Park, MD 20742, U.S.A.} \and Christopher Jarzynski\affil{2}{Department of Chemistry and Biochemistry, and Institute for Physical Science and Technology, University of Maryland, College Park, MD 20742, U.S.A.}}
	\contributor{Submitted to Proceedings of the National Academy of Sciences of the United States of America}

	\maketitle

	\begin{article}

	\begin{abstract}
We describe a minimal model of an autonomous Maxwell demon, a device that delivers work by rectifying thermal fluctuations while simultaneously writing information to a memory register.  We solve exactly for the steady-state behavior of our model, and we construct its phase diagram.  We find that our device can also act as a ``Landauer eraser'', using externally supplied work to remove information from the memory register.  By exposing an explicit, transparent mechanism of operation, our model offers a simple paradigm for investigating the thermodynamics of information processing by small systems.
	\end{abstract}

			
	\dropcap{A} system in thermal equilibrium undergoes random microscopic fluctuations, and it is tempting to speculate that an ingeniously designed device could deliver useful work by rectifying these fluctuations.  The suspicion that this would violate the second law of thermodynamics has inspired nearly 150 years of provocative thought experiments \cite{Maxwell1871, Smoluchowski1912, Szilard1929,  Brillouin1951, Feynman1}, leading to discussions of the thermodynamic implications of information processing \cite{Landauer1961, Bennett1982, Bennett1985, Zurek1989, Bub2001, Leff2003, Maroney2009}.  Although both Maxwell~\cite{Maxwell1871} and Szilard~\cite{Szilard1929} famously took the rectifying agent to be an intelligent being, later analyses have explored the feasibility of a fully mechanical  ``demon''.
	There has emerged a kind of consensus, based largely on the works of Landauer~\cite{Landauer1961} and Bennett~\cite{Bennett1982, Bennett1985}, and independently Penrose~\cite{Penrose1970}, according to which a mechanical demon can indeed deliver work by rectifying fluctuations, but in doing so it gathers information that must be written to physical memory.
	The eventual erasure of this information carries a thermodynamic cost, no less than $k_BT\ln 2$ per bit ({\it Landauer's principle}), which eliminates any gains obtained from the rectification of fluctuations.

	The past few years have seen increased interest in the thermodynamics of information processing~\cite{Quan2006,Andrieux2008,Maruyama2009, Lambson2011, Hosoya2011, Granger2011}.
	Discussions of Maxwell's demon, Landauer's principle and related topics arise in contexts such as quantum information theory~\cite{delRio2011}, the synthesis of artificial nanoscale machines~\cite{Kay2007}, feedback control in microscopic systems~\cite{Kim2007, Sagawa2008, Sagawa2009, Cao2009, Sagawa2010,Toyabe2010, Ponmurugan2010, Horowitz2010, Horowitz2011,Abreu2011,Vaikuntanathan2011,Sagawa2012}, and single-photon cooling of atoms~\cite{Raizen2011}.
	Experiments have been performed with the explicit aim of testing theoretical predictions \cite{Toyabe2010}, including Landauer's principle \cite{Berut2012}.
	Moreover the consensus or ``favored explanation''~\cite{Norton2011} described above is widely but not universally accepted, as suspicions persist that it assigns an unwarranted thermodynamic significance to random data~\cite{Maroney2009,Norton2011, Earman1998, Earman1999,Hemmo2010}.
	
	In spite of this attention, the field still lacks a tangible example or model of a device that converts heat into work at the expense of writing information.
	Discussions are often framed around general principles rather than a particular instance,
	and the demon is typically described in generic terms, as a system capable of performing microscopic feedback control, but otherwise unspecified.
	In this paper we propose an explicit, solvable model of a system that behaves as a Maxwell demon.
	Our device, which extracts energy from a single thermal reservoir and delivers it to raise a mass against gravity, is fully autonomous -- it is neither manipulated by an external agent nor driven by an explicit thermodynamic force -- but in order to lift the mass the device requires a memory register to which it can write information.
	
	\begin{figure*}[tbp]
	\centering
	\includegraphics[trim = 0.6in 8.6in 0in 1in, scale = 1.0]{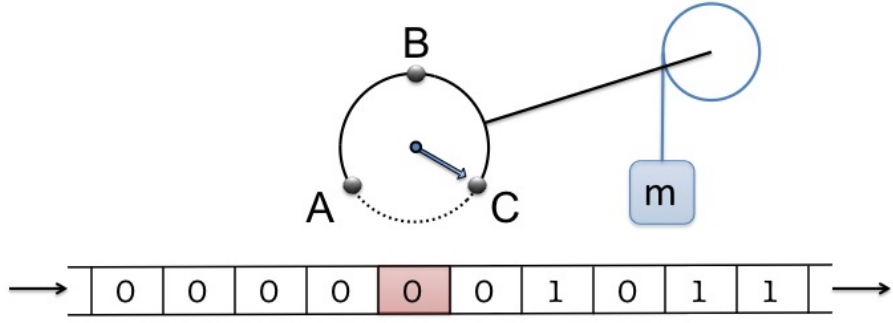}
	\caption{A sequence of bits moves at constant speed past the three-state demon, which interacts with the nearest bit (shaded) and a thermal reservoir (not shown).  To model a positive external load $f>0$, we imagine that a mass $m$ is lifted by an amount $\Delta h$ every time the demon makes a transition $C \rightarrow A$, and lowered with each transition $A \rightarrow C$.  See text for details.  For $f<0$, the mass can be pictured as hanging off the rights side of the small circle, so that transitions $C \rightarrow A$ lower the mass and transitions $A \rightarrow C$ lift it.}
	\label{fig:bit}
	\end{figure*}	

	Briefly, in our model the device, or demon, is a three-state system that interacts with: a thermal reservoir, a mass that can be lifted or lowered, and a stream of {\it bits} that pass by the demon in sequence, as sketched in Fig.~\ref{fig:bit}.
	The demon's dynamics consist of random transitions among its three states.
	These transitions are driven by thermal fluctuations from the reservoir, and are coupled to the bits and to the mass in a manner described in detail below.
	The model has three parameters: $\delta$ describes the initial statistical state of the bits, reflecting the initial ratio of {\tt 0}'s to {\tt 1}'s; $\epsilon$ characterizes the weight of the mass; and $\tau$ is the duration of interaction with each bit in the stream.
	For any set of values $(\delta,\epsilon,\tau)$ the model reaches a unique periodic steady state, characterized by an average rate of work performed on the mass, and an average rate of information written to the bit stream.

	We will first consider our model in the absence of an external load: the demon generates directed motion while writing information to the bits, but there is no provision for harnessing this motion to perform work.  We then add a load by attaching a non-zero mass, as in Fig.~\ref{fig:bit}, and we solve for the model's steady-state behavior.
	Our model exposes a specific mechanism for the operation of a mechanical Maxwell demon, allowing us to explore in detail the interplay between the gravitational pull on the mass and the changing information content of the stream of bits.
	Moreover, our demon is versatile: it is equally capable of acting as an eraser, using the energy of a falling mass to remove  information from the memory register.
	
\begin{figure*}[tbp]
	\centering
	\includegraphics[trim = 0.8in 7.8in 0in 1.3in , scale=1.0]{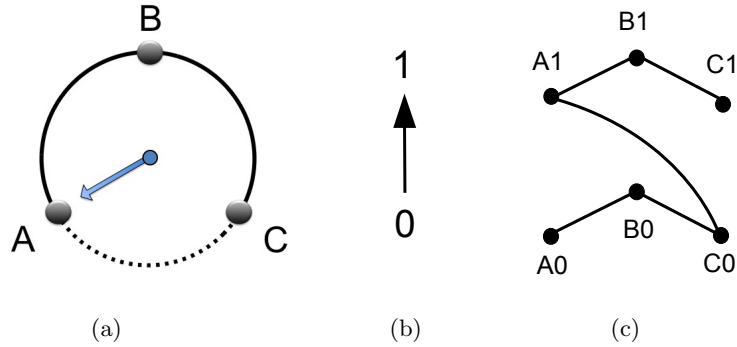}
	\caption{Schematic depiction of the demon, the bit and their composite 6-state system. (a) The state of the demon is indicated by an arrow pointing in one of three directions ($A$, $B$, or $C$) on the face of a dial. (b) The bit is represented as an arrow pointing either up ({\tt 1}) or down ({\tt 0}). (c) Network depiction of the composite system, showing allowed transitions.
	The edge that connects $A{\tt 1}$ and $C{\tt 0}$ represents the coupling between the demon and bit.}
	\label{fig:6state}
	\end{figure*}

	We now specify our model in detail, by introducing in turn its several elements (Fig.\ \ref{fig:6state}). The demon evolves by making thermally activated transitions among its three states, labelled $A$, $B$ and $C$.  We consider transitions in the direction $A \rightarrow B \rightarrow C \rightarrow A$ to be clockwise (CW), and those in the opposite direction to be counterclockwise (CCW), see Fig.\ \ref{fig:6state}(a).
The demon exhibits {\it directed rotation} if CW transitions occur with greater frequency than CCW transitions, or vice-versa.
	(Note that the demon must possess at least three states in order to exhibit directed rotation.)
	To keep track of the net CW rotation, we introduce an integer variable $\chi$, whose value increases by one unit whenever the demon makes a transition from $C$ to $A$, and decreases by one unit with each transition from $A$ to $C$.

	We next describe the interaction between the demon and a single bit, with states labelled {\tt 0} and {\tt 1}, Fig.\ \ref{fig:6state}(b). The demon and bit together form a composite system with six states, $A{\tt 0}, \cdots C{\tt 1}$, depicted in Fig. \ref{fig:6state}(c). The five lines, or edges, connecting pairs of states in this network specify the allowed transitions for the composite system. The demon can jump between states $A$ and $B$, and between $B$ and $C$, without involving the bit; these transitions are represented by the edges $A{\tt 0}$-$B{\tt 0}$, $B{\tt 0}$-$C{\tt 0}$, $A{\tt 1}$-$B{\tt 1}$ and $B{\tt 1}$-$C{\tt 1}$. Additionally, the demon can make a transition from $C$ to $A$ if the bit simultaneously ``flips'' from {\tt 0} to {\tt 1}, or from $A$ to $C$ if the bit flips from {\tt 1} to {\tt 0}, as indicated by the line connecting $A{\tt 1}$ and $C{\tt 0}$. We model these transitions as a Poisson process, where $R_{ij}$ is the probability per unit time to make a transition to state $i$, when the system is in state $j$, with $i,j \in \{ A{\tt 0}, \cdots C{\tt 1} \}$. For the moment, we set $R_{ij}=1$ for each of the ten allowed transitions (two per edge) depicted in Fig. \ref{fig:6state}(c).
	These rates set the unit of time in our model.  Because $R_{ij}=R_{ji}$ for every edge, we have implicitly assigned the same energy to all six states~\cite{vanKampen_3rdEd2007}.
	Under these rates, the demon and bit relax toward equilibrium, in which all six states are equally likely.
	This relaxation occurs on a time scale of order unity: $\tau_r \sim 1$.

	The network of states in Fig. \ref{fig:6state}(c) forms a linear chain. 
	Because this chain contains no closed loops, the model cannot yet exhibit directed rotation, only back-and-forth excursions along the chain.
To introduce the possibility of rotation, let us now imagine a {\it bit stream} (see Fig.\ \ref{fig:bit}): a sequence of bits arranged at equally spaced intervals along a tape that is pulled through at constant speed, for instance by a frictionless flywheel. The demon remains at a fixed location, and interacts, in the manner discussed above, only with the bit that is currently closest to it. Let $\tau^{-1}$ denote the rate at which the bits pass by the demon, each interacting with the demon for a time interval of duration $\tau$ before the next bit in the stream takes its place.
Thus $\tau$ determines the extent to which the composite system approaches equilibrium during one such {\it interaction interval}; for $\tau \ll \tau_r \sim 1$ the system hardly evolves during the interval, whereas for $\tau \gg 1$ the demon and the bit effectively reach equilibrium.
Finally, let ${\tt b}_n$ and ${\tt b}_n^\prime$ denote, respectively, the incoming and outgoing state of the $n$'th bit in the stream.  The state of any bit can change only when it is interacting with the demon.

	The incoming bits are statistically independent of one another, each with probability $p_0$ to be in state {\tt 0}, and $p_1$ to be in state {\tt 1}.
The {\it excess parameter}
\begin{equation}
\delta \equiv p_0 - p_1
\end{equation}
quantifies the excess of {\tt 0}'s in the incoming bit stream.

To understand how the demon can exhibit directed rotation, in this paragraph let us consider the case in which every bit in the incoming stream is set to {\tt 0} (as in Fig.~\ref{fig:bit}).
The demon interacts with the $n$'th bit during the $n$'th interaction interval, $t_n \le t < t_{n+1} \equiv t_n+\tau$.
At the start of this interval, the composite system begins in state $A{\tt 0}$, $B{\tt 0}$ or $C{\tt 0}$, since ${\tt b}_n={\tt 0}$.
From $t=t_n$ to $t_{n+1}$ the system evolves among the network of states depicted in Fig. \ref{fig:6state}(c).
It might repeatedly pass forward and back along the edge connecting $C{\tt 0}$ to $A{\tt 1}$,
resulting in alternating increments and decrements of the counter $\chi(t)$.
At the end of the interaction interval, if the system is found in state $A{\tt 0}$, $B{\tt 0}$ or $C{\tt 0}$ (i.e.\ if ${\tt b}_n^\prime={\tt 0}$) then we can infer that every transition $C{\tt 0} \rightarrow A{\tt 1}$ was balanced by a transition $A{\tt 1} \rightarrow C{\tt 0}$, hence $\Delta \chi_n \equiv \chi(t_{n+1}) - \chi(t_n) = 0$.
If the system instead ends in $A{\tt 1}$, $B{\tt 1}$ or $C{\tt 1}$ (${\tt b}_n^\prime={\tt 1}$), then the counter has advanced by one net unit: $\Delta \chi_n = +1$.
At $t=t_{n+1}$, the $n$'th bit is replaced by the $(n+1)$'th bit, and the next interval commences.
Thus if the composite system is in state $B{\tt 1}$ at the end of one interval, then at the start of the next interval it is in state $B{\tt 0}$.
This effective transition does not imply an actual change in the state of a given bit, but simply reflects the replacement of an outgoing bit in state {\tt 1} with an incoming bit in state {\tt 0}.
(Note that Fig. \ref{fig:6state}(c) depicts only the actual transitions that may occur during one interaction interval.)
Over time, the demon interacts with a sequence of bits, all initialized to {\tt 0}, and the outgoing bit stream contains a record of the demon's rotary motion: each occurrence of an outgoing bit in state {\tt 1} indicates one full CW rotation, $\Delta \chi = +1$.
Since the value of the counter can only increase or remain unchanged from one interval to the next, in the long run $\chi(t)$ grows with time and the demon undergoes directed CW rotation.

If the incoming stream were instead composed entirely of {\tt 1}'s, then full CW rotations would be prohibited, and full CCW rotations would be documented as outgoing {\tt 0}'s.
For a more general distribution of incoming bits, the net change in the counter during the $n$'th interaction interval is
\begin{equation}
\label{eq:deltachi}
\Delta \chi_n = {\tt b}_n^\prime - {\tt b}_n ,
\end{equation}
and the outgoing stream provides partial information regarding the demon's gyrations.

The demon eventually reaches a periodic steady state in which its statistical behavior is the same from one interval to the next.
If the outgoing bit stream is then characterized by values $p_0^\prime$, $p_1^\prime$ and $\delta^\prime \equiv p_0^\prime - p_1^\prime$, then the average number of full CW rotations per interaction interval is\begin{equation}
\label{eq:Phi}
\Phi \equiv \left\langle \Delta\chi_n \right\rangle = p_1^\prime - p_1 = \frac{1}{2}(\delta - \delta^\prime).
\end{equation}
We will use $\Phi$ as our measure of directed rotation, and we will call it the {\it circulation}.

We have solved for the periodic steady state, obtaining (as described in the Methods section below)
	\begin{subequations}
	\label{eq:soln_f=0}
	\begin{equation}
	\label{eq:Phi_f=0}
	\Phi(\delta;\tau) = \frac{\delta}{2}  \left[ 1 - \frac{1}{3}K(\tau) \right] ,
	\end{equation}
where
	\begin{equation}
	K(\tau) =
	e^{-2\tau} \frac{(1+8\alpha+4\sqrt{3}\beta) - (2+7\alpha+4\sqrt{3}\beta) e^{-2\tau}}{3 - (2+\alpha)e^{-2\tau}}
	\end{equation}
	\end{subequations}
and 
$\alpha = \cosh(\sqrt{3}\tau)$, $\beta = \sinh(\sqrt{3}\tau)$.
The function $K(\tau)$ decreases monotonically from $K(0^+)=3$ to $K(\infty)=0$, hence the magnitude of the circulation increases with $\tau$, from $\Phi(\delta;0^+)=0$ to $\Phi(\delta;\infty) = \delta /2$.
These values are easily understood.
When $\tau\rightarrow 0$, the probability to observe any transition during a given interaction interval vanishes, and therefore so does $\Phi$.
When $\tau\gg 1$, during each interaction interval the composite system has sufficient time to relax to equilibrium, with all six states in Fig. \ref{fig:6state}(c) equally likely, hence $\delta^\prime = 0$ and $\Phi = \delta/2$ (Eq.~\ref{eq:Phi}).
Interestingly, the average {\it rate} of rotations, $\omega=\Phi/\tau$, achieves its maximal value $\omega=\delta /6$ at $\tau=0^+$, and then decreases with $\tau$.
	
	Since the quantity in square brackets in Eq.~\ref{eq:Phi_f=0} is non-negative, the demon effectively converts an excess of {\tt 0}'s or {\tt 1}'s into directed CW or CCW rotation.
	We now harness this rotation to an external load, by attaching a mass $m$ to the demon in such a way that the mass is lifted a distance $\Delta h$ whenever the demon makes a transition from $C$ to $A$, and is lowered by $\Delta h$ for the reverse transition (Fig.~\ref{fig:bit}).
	The corresponding energy, $\pm mg\Delta h$, is exchanged with the thermal reservoir: with every transition $C{\tt 0}\rightarrow A{\tt 1}$, heat is withdrawn from the bath to lift the mass, and with every transition $A{\tt 1}\rightarrow C{\tt 0}$ that energy is released to the bath.
	(There is no exchange of energy with the flywheel that pulls the bits past the demon, in particular the flywheel does not contribute energy to lift the mass.)
	To incorporate these considerations into our dynamics, we modify the transition rates in accordance with detailed balance~\cite{vanKampen_3rdEd2007}:
	\begin{equation}
	\label{eq:ratio}
	\frac{R_{A1,C0}}{R_{C0,A1}} = e^{-f}
	\quad,\quad
	f \equiv\frac{mg\Delta h}{k_BT} > 0,
	\end{equation}
	where $T$ is the temperature of the thermal reservoir and $k_B$ is Boltzmann's constant, with all other rates $R_{ij}$ unchanged.
	The parameter $f$ quantifies a thermodynamic force that favors CCW rotations ($R_{C0,A1}>R_{A1,C0}$) as gravity tugs the mass downward.
	In terms of Fig.~\ref{fig:6state}(c), Eq.~\ref{eq:ratio} effectively increases the energies of states $A{\tt 1}$, $B{\tt 1}$ and $C{\tt 1}$ by an amount $mg\Delta h$, relative to states $A{\tt 0}$, $B{\tt 0}$ and $C{\tt 0}$, reflecting the energy that is withdrawn from the reservoir during the transition $C{\tt 0}\rightarrow A{\tt 1}$.

	If the demon interacts with a single bit for a sufficiently long time then the two will reach  equilibrium, with
	\begin{subequations}
	\label{eq:equilibrium}	
	\begin{equation}
	p_{A0}^{\rm eq} = p_{B0}^{\rm eq} = p_{C0}^{\rm eq} = \frac{e^{f}}{Z} \quad , \quad
	p_{A1}^{\rm eq} = p_{B1}^{\rm eq} = p_{C1}^{\rm eq} = \frac{1}{Z},
	\end{equation}
where $Z=3(1+e^{f})$.
	After summing over the states of the demon, the equilibrium probabilities for the bit itself are found to satisfy
	\begin{equation}
	\label{eq:equilibrium_bit}
	p_0^{\rm eq} - p_1^{\rm eq} = \tanh\left(\frac{f}{2}\right) \equiv \epsilon \quad ,
	\end{equation}
	\end{subequations}
	where the {\it weight parameter}, $\epsilon$, is a rescaled version of the thermodynamic force, $f$.
	
	Eq.~\ref{eq:ratio} leaves us with some freedom in assigning the rates $R_{C0,A1}$ and $R_{A1,C0}$.  We have chosen
	\begin{equation}
	R_{A1,C0} = 1 - \epsilon \quad , \quad R_{C0,A1} =  1 + \epsilon \quad.
	\end{equation}
	With this choice, we are again able to solve analytically for the periodic steady state, obtaining (see Methods)
	\begin{subequations}
	\label{eq:centralResult}
	\begin{equation}
	\label{eq:Phif}
	\Phi ( \delta, \epsilon; \tau) = \frac{ \delta - \epsilon }{2} \bigg [ 1 - \frac{1}{3} K( \tau ) + \frac{ \epsilon \delta}{6} J( \tau, \epsilon \delta) \bigg ]
	\end{equation}
	with
	\begin{eqnarray}
	\label{eq:J}
	J ( \tau, \epsilon \delta)  \hspace{2.75 in} \\ \nonumber
	 = \frac{(1-e^{-\tau}) [2 e^{-2 \tau} (\alpha + \sqrt{3} \beta - 1)]^2}{ [ 3 (1 - \epsilon \delta e^{-\tau}) - (1-\epsilon \delta)(2 + \alpha)e^{-2 \tau}] [3 - (2 + \alpha) e^{-2 \tau}]} 
	\end{eqnarray}
	\end{subequations}
and $K$, $\alpha$ and $\beta$ as in Eq. \ref{eq:soln_f=0}.
	These results extend to negative values of $f$, if we interpret these as indicating that gravity exerts a {\it clockwise} torque (see caption of Fig.~\ref{fig:bit}).
	Eq.~\ref{eq:centralResult}, our central result, is then valid for $\vert\epsilon\vert < 1$, $\vert\delta\vert\le 1$, and $0<\tau<\infty$.

	In the limits $\tau=0^+$ and $\tau\rightarrow\infty$ Eq.~\ref{eq:centralResult} gives $\Phi=0$ and $\Phi \rightarrow (\delta-\epsilon)/2$, respectively.
	The latter reflects equilibration between the demon and each bit: $p_{0,1}^\prime = p_{0,1}^{\rm eq}$, hence $\delta^\prime \equiv p_0^\prime - p_1^\prime = \epsilon$ (see Eqs.~\ref{eq:Phi} and \ref{eq:equilibrium_bit}).
	On the right side of Eq.~\ref{eq:Phif}, the quantity in square brackets is non-negative, as determined by numerical inspection, and the prefactor indicates a competition 
	between the parameters $\delta$ and $\epsilon$.
	When $\delta > \epsilon$, the incoming bits contain a surplus of ${\tt 0}$'s, relative to the equilibrium proportions (Eq.~\ref{eq:equilibrium}); during each interaction interval the composite system relaxes toward equilibrium, generating CW rotation as ${\tt 0}$'s are converted to ${\tt 1}$'s, on average.
	Similarly when $\delta < \epsilon$ the relative surplus of incoming ${\tt 1}$'s generates CCW rotation.
	When $\delta = \epsilon$ there is no directed rotation, as the bits arrive distributed in the equilibrium ratio.
		
	The new term appearing in Eq.\ \ref{eq:Phif},  $\epsilon \delta J/6$, does not affect the sign of the quantity in square brackets. However, there is a succinct way to describe its action: it goes against the loser if there is a competition between $\delta$ and $\epsilon$, and against both if there is cooperation.  This follows from the inequality $J( \tau, \epsilon \delta) \geq 0$ (determined by numerical inspection).
	E.g.\ if $\delta > \epsilon > 0$, so that in competition $\delta$ wins giving rise to $\Phi>0$, then the term $\epsilon \delta J/6$ makes a positive contribution to $\Phi$. If $\delta > 0 > \epsilon$, so that both parameters favor CW rotation, the contribution due to this term is negative.
		
	\begin{figure*}[tbp]
	\centering
	\includegraphics[trim = 0in 6.8in 0in 1.5in, scale = 0.91]{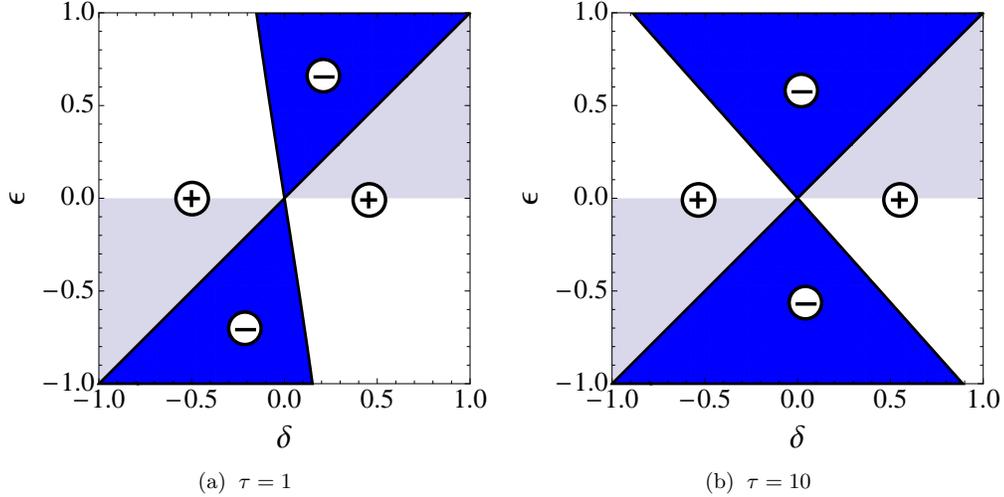}
	\caption{Behavior of our model as a function of $\delta$ and $\epsilon$, for $\tau=1$ and 10.
	The demon can act as an engine (lightly shaded region), an eraser (darkly shaded) or a dud (unshaded).
	These regions are delineated by the lines $\epsilon=0$ and $\epsilon=\delta$, together with a third line (see text), shown passing through the second and fourth quadrants, which depends on $\tau$ and is nearly but not exactly straight.
	The symbols $+$ and $-$ indicate the sign of $\Delta S$, the average change in disorder per bit.
	The circulation $\Phi$ is positive (CW) in the lower right half of the figure, and negative (CCW) in the upper left.}
	\label{fig:phaseDiagram}
	\end{figure*}
	
	We now explore the thermodynamic and information-processing behavior of our device, and we ask when it might perform a ``useful'' service.
	Consider the square region representing allowable values of the excess parameter $\delta$ and the weight parameter $\epsilon$, depicted in Fig.~\ref{fig:phaseDiagram} for $\tau=1$ and $10$.
	The line $\epsilon=\delta$ is the contour of zero steady-state rotation: to the left of this line the rotation is CCW ($\Phi<0)$ and to the right it is CW ($\Phi>0$).
	The product $W \equiv k_BTf\Phi$ represents the average work that the device delivers to the mass, per interaction interval.
	Since $\text{sign}(\epsilon)=\text{sign}(f)$, the two lightly shaded triangles in Fig.~\ref{fig:phaseDiagram} ($\epsilon\Phi>0$) are the regions in which the device acts as an {\it engine}, converting heat from the thermal bath into work to lift the mass.
	For example, when $\delta >\epsilon >0$ gravity exerts a CCW torque, but the excess of incoming {\tt 0}'s generates a greater CW torque.
	
	Now consider the quantities
	\begin{equation}
        \label{eq:Sb}
        S_b = - \sum_{i = 0, 1} p_i \ln{p_i}
        \quad \text{and} \quad
        S_b^\prime = - \sum_{i = 0, 1} p_i^\prime \ln{p_i}^\prime.
        \end{equation}
For convenience we will call these the {\it disorder} (per bit), although this terminology ignores correlations between successive bits in the outgoing stream.
	$S_b$ quantifies the information content of the incoming stream, and is related to its capacity to record new information, in the following sense.
	When $S_b=0$ the incoming stream is a blank slate composed entirely of {\tt 0}'s (or entirely of {\tt 1}'s), and the outgoing stream contains a faithful record of CW (or CCW) rotations, as discussed earlier.
	When $S_b=\ln 2$ (its maximum possible value) the incoming stream is saturated with an equal mixture of {\tt 0}'s and {\tt 1}'s, and in this case the outgoing stream does not chronicle the demon's rotations.
	We will interpret the difference $\Delta S \equiv S_b^\prime - S_b$ as a measure of the degree to which new information is written to the bits, as they interact with the demon.

	Since the rotation of the demon couples tightly to the flipping of bits (Eq.~\ref{eq:deltachi}), the line $\epsilon=\delta$ (where $\Phi=0$) is a contour along which $\Delta S(\delta,\epsilon;\tau)=0$; here, there is no net rotation and no net change in the bit statistics: $p_0^\prime=p_0$ and $p_1^\prime=p_1$.
	The other solid line depicted in Fig.~\ref{fig:phaseDiagram}, running from the upper left to the lower right, is also a contour along which $\Delta S = 0$, representing the {\it inversion} of bit statistics: $p_0^\prime = p_1$ and $p_1^\prime=p_0$.
	The two lines divide the $(\delta,\epsilon)$-square into four regions, with the $+$'s and $-$'s in Fig.~\ref{fig:phaseDiagram} denoting the sign of $\Delta S$ in these regions.
				
	We see in Fig.~\ref{fig:phaseDiagram} that $\Delta S > 0$ whenever our device acts as an engine.
	This is consistent with the proposition that a mechanical demon, in order to convert heat to work, must write information to a memory register.	
	Indeed, Fig.~\ref{fig:phaseDiagram} shows that the greater the storage capacity of the incoming bit stream, the larger the mass the demon can hoist against gravity:
	when presented with a blank slate ($\delta = \pm 1$) the demon can lift any mass; but when the incoming bit stream is saturated ($\delta=0$) the demon is incapable of delivering work.
	Thus a blank or partially blank memory register acts as a thermodynamic resource that gets consumed when the demon acts as an engine.

	In the above description, the demon is an active rectifying agent and the bit stream merely a passive receptacle for information.  From another perspective, however, the interaction with the demon presents an opportunity for the bits to evolve to a more disordered sequence of {\tt 0}'s and {\tt 1}'s.  The bits' role then appears more assertive: their evolution toward greater randomness is what drives the engine, and the demon simply facilitates the process.
	
	In the darkly shaded regions in Figs.~\ref{fig:phaseDiagram}, the demon acts as an {\it eraser}, removing information from the memory register: $\Delta S < 0$.
	For example, if $\delta=0$, $f\gg 1$ (i.e.\ $\epsilon\approx 1$) and $\tau\gg 1$, then the bits arrive in an equal mixture of {\tt 0}'s and {\tt 1}'s, but each bit has sufficient time to equilibrate with the demon, hence at the end of each interaction interval the composite system is almost certainly in state $A{\tt 0}$, $B{\tt 0}$ or $C{\tt 0}$ (Eq.~\ref{eq:equilibrium}).
	As a result, the outgoing bits are nearly all {\tt 0}'s, and the memory is effectively wiped clean as the mass drops by a distance $\Delta h/2$ (on average) per interaction interval.
		
	Our model thus reflects the interplay between two effective forces, one associated with the randomization of the bits and the other with the pull of gravity.	
	When our model acts as an engine, it consumes one resource -- a blank or partially blank memory register -- to build up another: the gravitational potential of the mass.  When it acts as an eraser the roles are reversed.
	In the unshaded regions in Figs.~\ref{fig:phaseDiagram}, both resources are squandered (the mass falls and the bits' disorder increases) and our model is a dud, accomplishing nothing useful.
	
	Our model satisfies the inequality
	\begin{equation}
	\label{eq:ineq}
	W \le k_BT\Delta S ,
	\end{equation}
	for any $\epsilon$, $\delta$ and $\tau$, with the equality holding only when $\epsilon=\delta$.
	(See Supplementary Material.)
Thus, the increase in the information content of the bit stream places an upper limit on the work that can be delivered, when the model is an engine.
Analogous inequalities arise in the context of feedback control, where an external agent manipulates the system on the basis of outcomes of explicit measurements~\cite{Kim2007, Sagawa2008, Sagawa2009, Cao2009, Sagawa2010,Toyabe2010, Ponmurugan2010, Horowitz2010, Horowitz2011,Abreu2011,Vaikuntanathan2011,Sagawa2012}.
When our model acts an eraser ($\Delta S < 0$), Eq.~\ref{eq:ineq} reveals the minimum amount of work that must be {\it supplied}, by the falling mass, in order to reduce the information content by a given amount.
In the case of full erasure ($S_b^\prime=0$) this becomes Landauer's principle, $\vert W\vert > k_B TS_b$.
Note that if we are willing to assign thermodynamic meaning to the randomness in a string of data, Eq.~\ref{eq:ineq} can be interpreted as the second law of thermodynamics (or rather as a weak statement of it, since $S_b^\prime$ ignores correlations between outgoing bits): the decrease in the entropy of the reservoir, $- \Delta S_r = W/k_BT$, must not exceed the increase in the entropy of the bit stream:
	\begin{equation}
	\label{eq:2nd_law}
	\Delta S_r + \Delta S \geq 0.
	\end{equation}
	
	While both sides of Eq.~\ref{eq:ineq} approach zero as $\epsilon\rightarrow\delta$, their ratio approaches unity in that limit (see Supplementary Material).
	Thus in the immediate vicinity of the line $\epsilon=\delta$, the bound represented by Eq.~\ref{eq:ineq} becomes saturated, and our model behaves with maximal efficiency, acting as a thermodynamically {\it reversible} engine or eraser.
	Note however that the {\it rate} at which the demon either delivers work or erases information approaches zero in this reversible limit.
	
	We conclude by mentioning two extensions of the present work.
	First, we can reformulate our model as one in which the bits arrive as a sequence of matched pairs, or {\it dimers}, ${\tt b}_{2k-1} = {\tt b}_{2k}$	(e.g.\ ${\tt 00}\, {\tt 11}\, {\tt 11}\, {\tt 00}\, {\tt 11} \cdots$), and the demon interacts with the bit stream, one dimer at a time.
	Even if the incoming stream contains an equal proportion of {\tt 0}'s and {\tt 1}'s, the demon is able to lift the mass, effectively by ``digesting'' the pairwise correlations between the bits, which depart in a less ordered sequence (e.g.\ ${\tt 01}\, {\tt 11}\, {\tt 10}\, {\tt 10}\, {\tt 11} \cdots$).
	This suggests the possibility of a more complex information-processing engine, driven by the recognition of specific patterns in the bit stream.
	
	We have also sketched a mechanistic version of our model, composed of frictionless paddles, pulleys and axles immersed in a dilute gas.
	While highly idealized, this model is more easily visualized as a material physical system than the discrete-state model described in the present paper.
	See Ref.~\cite{Bennett1985} for an analogous model of a Turing machine.

	\section{Methods}
	\label{sec:methods}
	
	To obtain $\Phi(\delta, \epsilon; \tau)$ we solve for the periodic steady state of the demon, then we use that solution to determine the distribution of outgoing bits $(p_0^\prime,p_1^\prime)$.
	The value of $\Phi$ then follows from Eq.~\ref{eq:Phi}.
	
	Let $\mathcal{T}_{3 \times 3}$ denote the transition matrix whose component $T_{\mu\nu}$ gives the probability to find the demon in state $\mu\in\{A,B,C\}$ at the end of one interaction interval, given that it began in state $\nu$ at the start of the interval.
	Explicitly,
	\begin{equation}
	\mathcal{T} = \mathcal{P}_D e^{\mathcal{R} \tau} \mathcal{M}.
	\end{equation}
Here, $\mathcal{M}_{6 \times 3} = \left ( \begin{array}{c}
										p_0 \, \mathbb{I} \\
										p_1 \, \mathbb{I} 
										\end{array} \right )$, with $\mathbb{I}$ the $3\times 3$ identity matrix; $\mathcal{R}_{6\times 6}$ is the transition rate matrix whose elements $\{R_{ij}\}$ are discussed in the main text; and $\left ( \mathcal{P}_D \right )_{3 \times 6} = \left ( \mathbb{I}, \mathbb{I} \right )$.
Specifically, if ${\bf q}_0 \equiv(q_0^A,q_0^B,q_0^C)^T$ gives the probability distribution that describes the demon at the start of one interval, then the six-component vector $\mathcal{M}{\bf q}_0$ gives the combined state of the (initially uncorrelated) demon and bit.
The factor $e^{\mathcal{R}\tau}$ propagates this distribution for the duration of the interval, and $\mathcal{P}_D$ projects out the state of the bit, so that ${\bf q}_\tau = \mathcal{T}{\bf q}_0$ is the statistical state of the demon at the end of the interval.

	In the periodic steady state, the initial probability distribution of the demon is given by the vector satisfying $\mathcal{T} \mathbf{q}^{\rm pss}=\mathbf{q}^{\rm pss}$, whose uniqueness is guaranteed by the Perron-Frobenius Theorem~\cite{Mayer2000}.
	At the end of the interaction interval the state of the (now correlated) demon and bit is $e^{\mathcal{R} \tau} \mathcal{M} \mathbf{q}^{\rm pss}$.
	The statistics of the outgoing bit are obtained by projecting this correlated state to that of the bit:
\begin{equation}
\label{eq:outgoingBits}
\left(
\begin{array}{c} p_0' \\ p_1'
\end{array}
\right)
= \mathcal{P}_B e^{\mathcal{R} \tau} \mathcal{M} \mathbf{q}^{\rm pss} \quad , \quad 
\mathcal{P}_B \equiv \left ( \begin{array}{cccccc}
																	1 & 1 & 1 & 0 & 0 & 0 \\
																	0 & 0 & 0 & 1 & 1 & 1
																	\end{array} \right ).
\end{equation}

Thus to obtain $(p_0^\prime,p_1^\prime)$ we must determine $\mathcal{T}$, solve for its eigenstate $\mathbf{q}^{\rm pss}$, and apply Eq.~\ref{eq:outgoingBits}.
This calculation involves a straightforward if tedious exercise in the spectral decomposition of $\mathcal{R}$, which we detail in the Supplementary Material.
%




\begin{acknowledgments}
	We thank Andy Ballard, J.\ Robert Dorfman, Jordan M.\ Horowitz, Juan M.\ R.\ Parrondo, Haitao Quan and Suriyanarayanan Vaikuntanathan for stimulating discussions, and we gratefully acknowledge financial support from the National Science Foundation (USA) under grants DMR-0906601 and ECCS-0925365, and the University of Maryland, College Park.

\end{acknowledgments}



\end{article}

	
	\section{Supplementary Material}
	\label{sec:supplementary}
	
	Here we detail key steps in the derivation of our expression for $\Phi(\delta, \epsilon; \tau)$, and we discuss the inequality $W \le k_BT \Delta S$, Eq.\ 10 of the main text.
	
	\subsection{Solving for $\Phi$}
	
	As explained in the ``Methods" section, to solve for $\Phi$ we first obtain the stationary probability distribution $\mathbf{q}^{pss}$ of the transition matrix $\mathcal{T}_{3\times 3}$.  This matrix describes the evolution of the demon over one interaction interval, and is the product of three matrices, $\mathcal{T} = \mathcal{P}_D e^{\mathcal{R} \tau} \mathcal{M}$ (Eq.\ 12).  Expressions for $\mathcal{P}_D$ and $\mathcal{M}$ were provided explicitly in the ``Methods'' section.  For $\epsilon \in (-1,1)$, the transition rate matrix for the composite demon-and-bit is
	\begin{equation}
	\label{eq:Re}
	\mathcal{R} = \left(
		\begin{matrix}
		-1 & 1 & 0 & 0 & 0 & 0 \\
		1 & -2 & 1 & 0 & 0 & 0 \\
		0 & 1 & -2+\epsilon & 1+\epsilon & 0 & 0 \\
		0 & 0 & 1-\epsilon & -2 - \epsilon & 1 & 0 \\
		0 & 0 & 0& 1 & -2 & 1 \\
		0 & 0 & 0 & 0 & 1 & -1 
		\end{matrix} \right).
	\end{equation}
This matrix has six real, non-degenerate eigenvalues that are (surprisingly) independent of $\epsilon$:
\begin{equation}
\label{eq:eigvals}
 \{ \lambda_i \} = \{ 0, -c, -1, -2, -3, -d \}
 \quad,
 \end{equation}
 with
 \begin{equation}
\begin{split}
a &= 1 - \sqrt{3} \quad,\quad
c = 2 - \sqrt{3} \quad,\quad x = 1+\epsilon \\
b &= 1 + \sqrt{3} \quad,\quad
d = 2 + \sqrt{3} \quad,\quad y = 1-\epsilon .
\end{split}
\end{equation}
The quantities $a$, $b$, $x$ and $y$ will be used momentarily.
 
 We have found the following spectral decomposition of $\mathcal{R}$ to be convenient:
 \begin{equation}
 \begin{split}
 \mathcal{R} 
&= \sum_{i = 1}^6 \frac{| i \rangle \lambda_i \langle i |}{\langle i | i \rangle}
 =  U N^{-1} \Lambda V \\
&=
\left(
\begin{array}{ccc}
\uparrow && \uparrow \\
{\bf u}_1 &\cdots& {\bf u}_6 \\
\downarrow && \downarrow
\end{array}
\right)
\left(
\begin{array}{ccc}
n_1^{-1} && \\
& \ddots & \\
&& n_6^{-1}
\end{array}
\right)
\left(
\begin{array}{ccc}
\lambda_1 && \\
& \ddots & \\
&& \lambda_6
\end{array}
\right)
\left(
\begin{array}{ccc}
\leftarrow & {\bf v}_1 & \rightarrow \\
& \vdots & \\
\leftarrow & {\bf v}_6 & \rightarrow
\end{array}
\right).
\end{split}
\end{equation}
Here, the columns of $U$ are right eigenvectors of $\mathcal{R}$, and the rows of $V$ are its left eigenvectors.
We denote the right eigenvectors by ${\bf u}_i$ or $\vert i\rangle$, and the left eigenvectors by ${\bf v}_i^T$ or $\langle i \vert$.
These form a biorthogonal pair of basis sets: ${\bf v}_i^T \cdot {\bf u}_j = \langle i\vert j\rangle = n_i \delta_{ij}$, i.e.\ $VU=N$.
Explicitly,
	\begin{equation}
	U = \left(
		\begin{matrix}
		x & 1 & x &1 & x & 1 \\
		x & -a & 0 & -1 & -2x & -b \\
		x & c & -x & -1 & x & d \\
		y & -c & -y & 1 & y & -d \\
		y & a & 0 & 1 & -2y & b \\
		y & -1 & y & -1 & y & -1
		\end{matrix} \right)
		,\quad
	V =  \left(	\begin{matrix}
								1 & 1 & 1 & 1 & 1 & 1\\
								y & -a y & c y & - c x & a x & -x\\
								1 & 0 & -1 & -1 & 0 & 1\\
								y & -y & -y & x & x & -x \\
								1 & -2 & 1 & 1 & -2 & 1\\
								y & -b y & d y & - d x & b x & -x
								\end{matrix} \right),
	\end{equation}
and $\{ n_i\} = \{ 6, 12 c, 4, 6, 12, 12 d\}$.
Note that since $\mathcal{R}$ is not symmetric, its left and right eigenvectors differ.
The matrices $N$ and $\Lambda$ are diagonal.
While it is usual to normalize the left and right eigenvectors so that they are biorthonormal ($n_i=1$), we have found that the choice of normalization given above leads to less cumbersome expressions in the subsequent analysis.
 
 In terms of this decomposition, we have
	\begin{equation}
	\mathcal{T} = \mathcal{P}_D e^{\mathcal{R} \tau} \mathcal{M}
	  =  \Bigl( \mathbb{I} \quad \mathbb{I}  \Bigr)
	   U N^{-1} e^{\Lambda\tau} V\left ( \begin{array}{c}
										p_0 \, \mathbb{I} \\
										p_1 \, \mathbb{I} 
										\end{array} \right )
	\end{equation}
	where $\mathbb{I}$ is the $3\times 3$ identity matrix (see ``Methods'').
An explicit evaluation yields
\begin{equation}
\begin{split}
	\mathcal{T} &= 
	\frac{1}{12}
	\left ( \begin{array}{ccc}
		F + G + \delta H \qquad & M - 2\delta L \qquad & F - G + \delta H \\
		M\qquad & M + 12 \sigma^3\qquad & M \\
		F - G - \delta H \qquad & M + 2\delta L \qquad & F + G - \delta H
										\end{array} \right ) \\
	&- 
	\frac{\epsilon}{12}
	\left ( \begin{array}{ccc}
		H + \delta (G-6\sigma) \qquad & - 2 L \qquad & H - \delta (G-6\sigma) \\
		0\qquad & 0 \qquad & 0 \\
		- H - \delta (G-6\sigma) \qquad &  2 L \qquad & - H + \delta (G-6\sigma)
										\end{array} \right )
\end{split}
\end{equation}
where $\sigma = e^{-\tau}$ and
\begin{equation}
\begin{split}
F = 4 + 2\sigma^3 \qquad,\qquad
G = 4 \sigma^2 &+ \sigma^c + \sigma^d \qquad,\qquad
H = \sqrt{3}(\sigma^c - \sigma^d) \\
L = 2 \sigma^2 - \sigma^c - \sigma^d \qquad&,\qquad
M = 4 - 4 \sigma^3
\end{split}
\quad.
\end{equation}

\vspace{.5in}

Solving the equation $\mathcal{T} {\bf q}^{\rm pss} = {\bf q}^{\rm pss}$ (see ``Methods'') we obtain
\begin{equation}
	{\bf q}^{\rm pss} = \frac{1}{3}
	\left( \begin{array}{c}
		1 + N \\ 1 \\ 1 - N
	\end{array}\right)
	\qquad,\qquad
	N(\delta,\epsilon) = \frac{(\delta-\epsilon)(H-L)}{6-G+\epsilon\delta(G-6\sigma)}
	\quad.
\end{equation}
Combining this result with Eq.\ 13 of the text yields the statistics of the outgoing bits, $(p_0^\prime,p_1^\prime)$, from which we then obtain the circulation using the relation $\Phi = p_1^\prime - p_1$.
		
	\subsection{Relationship between $W$ and $\Delta S$}
									
	We now obtain Eq.\ 10 of the main text: $W \leq k_B T \Delta S$. Since
	\begin{equation}
	W = k_B T \Phi  f = \, k_B T \Phi \ln \frac{1 + \epsilon}{1-\epsilon}
	,
	\end{equation}
	we must establish the non-negativity of the \emph{dissipation} function:
	\begin{equation}
	\label{eq:Wd}
	\Omega \equiv \Delta S - \Phi \, \ln{\frac{1 + \epsilon}{1-\epsilon}} \geq 0.
	\end{equation}
We will first prove this for the quasistatic case $\tau \rightarrow \infty$, and then extend it to finite $\tau$.  Let
	\begin{equation}
	\label{eq:entropy}
	S(X) = - \frac{1 - X}{2} \ln{\frac{1 - X}{2}} - \frac{1 + X}{2} \ln{\frac{1 + X}{2}}
	\end{equation}
denote the entropy of a bit as a function of an excess parameter $X$.

In the quasistatic limit (specified below by the subscript $``\infty"$) the outgoing bits reflect full equilibration between the demon and the bit; see Eq.\ 6 of the main text.  In that limit we have
	\begin{equation}
	\label{eq:limitcomp}
	\Phi  \longrightarrow  \frac{\delta - \epsilon}{2} \equiv \Phi_{\infty} \quad,\quad
	\delta' \longrightarrow  \epsilon  \quad,\quad
	\Delta S_b  \longrightarrow   S(\epsilon) - S(\delta),
	\end{equation}
hence
	\begin{equation}
	\label{eq:limit}
	\Omega  \longrightarrow  S(\epsilon) - S(\delta) - \frac{\delta - \epsilon}{2} \ln{\frac{1 + \epsilon}{1 - \epsilon}} \equiv \Omega_{\infty}.
	\end{equation}
	Now note that
	\begin{equation}
	\Omega_{\infty}  = 0 \quad \text{for $\epsilon = \delta$}
\qquad \text{and} \qquad
	\frac{\partial}{\partial\epsilon} \Omega_{\infty} = \frac{\epsilon - \delta}{1 - \epsilon^2}	 \left \{	\begin{array}{rl}
																	> 0 & \text{if $ \epsilon > \delta$}\\
																	< 0 & \text{if $\epsilon < \delta$}
																	\end{array} \right.
																	\quad.
	\end{equation}
	Thus for any fixed value of $\delta$, the function $\Omega_{\infty}(\delta,\epsilon)$ is zero at the point $\epsilon=\delta$, and as a function of $\epsilon$ it decreases when $\epsilon < \delta$ and increases when $\epsilon>\delta$.  This establishes that $\Omega_{\infty} \ge 0$.
	
	We have verified by explicit numerical investigation that 
	\begin{equation}
	\label{eq:eta}
	\Phi(\delta, \epsilon; \tau) = \eta \, \Phi_{\infty} \quad \text{where} \quad 0 \leq \eta \leq 1.
	\end{equation}
That is, the quantity in square brackets in Eq.\ 8a of the main text (which we here label $\eta$) falls in the range $[0,1]$.
While we have not been able to establish this analytically, we believe it is related to the fact that all eigenvalues of the transition rate matrix $\mathcal{R}$ are real and non-positive (Eq.~\ref{eq:eigvals}), with the consequence that the composite demon-and-bit system relaxes monotonically toward equilibrium during each interaction interval.
This suggests that $\text{sign}(\Phi) = \text{sign}(\Phi_\infty)$, and that the maximum circulation is obtained by allowing the composite system to relax fully to equilibrium.
	
	For finite $\tau$, the excess parameter $\delta^\prime$ for the outgoing stream is a linear average of $\delta$ and $\epsilon$:
	\begin{equation}
	\delta^\prime =  \delta - 2 \Phi = (1 - \eta) \, \delta + \eta \, \epsilon \quad,
	\end{equation}
	using Eq. 3 of the main text, Eq.~\ref{eq:eta} and $\Phi_\infty = (\delta-\epsilon)/2$. Since $S(X)$ is concave (${\rm d}^2S/{\rm d}X^2<0$),
	\begin{equation}
	\label{eq:tau1}
	\begin{array}{ll}
	S_b' = S(\delta') & \geq (1-\eta) \, S(\delta) + \eta \, S(\epsilon)\\
	& = S(\delta) + \eta \, \big [ S(\epsilon) - S(\delta) \big]. 
	\end{array}
	\end{equation}
	From Eqs.~\ref{eq:limitcomp}, \ref{eq:limit} and the non-negativity of $\Omega_{\infty}$, we have
	\begin{equation}
	\label{eq:tau2}
	S(\epsilon) - S(\delta) \geq \Phi_{\infty} \ln{\frac{1 + \epsilon}{1 - \epsilon}}.
	\end{equation} 
Combining Eqs.~\ref{eq:tau1} and \ref{eq:tau2} we get
	\begin{equation}
	S_b' \geq S(\delta) + \eta \, \Phi_\infty \ln{\frac{1 + \epsilon}{1 - \epsilon}}
	= S_b + \Phi \ln{\frac{1 + \epsilon}{1 - \epsilon}} \quad,
	\end{equation}
	which is the result we set out to establish (Eq.~\ref{eq:Wd}).
	
	As mentioned in the main text, the inequality $W \le k_BT \Delta S$ can be viewed as a weak statement of the second law of thermodynamics, where the weakness is due to the neglect of correlations in the outgoing bit stream.\footnote{
	Empirically, these correlations are quite small, though non-zero.}
	If we accept an equivalence between thermodynamic entropy and the information content of a random data set, then we might expect the second law to be represented more accurately by the inequality
	\begin{equation}
	\label{eq:strong}
	W \leq k_B T \Delta \mathcal{H} \quad,
	\end{equation}
	where $\Delta \mathcal{H} = \mathcal{H}_b^\prime - \mathcal{H}_b$ is the change in the information entropy per bit, {\it including} correlations among the outgoing bits.
	For our model Eq.~\ref{eq:strong} can be derived directly from general properties of Markov processes, without involving our solution for $\Phi(\delta,\epsilon;\tau)$.
	We omit this derivation, which makes use of relative entropy\footnote{Cover T M, Thomas J A (2006) {\it Elements of Information Theory} (Wiley-Interscience, Hoboken, New Jersey).} as a Lyapunov function\footnote{Schnakenberg J (1976) {\it Network Theory of Microscopic and Macroscopic Behavior of Master Equation Syatems}. Rev Mod Phys 48:571-585.} characterizing the relaxation of the demon and bits.
	Since $\mathcal{H}_b = S_b$ (the bits arrive uncorrelated) and $\mathcal{H}_b^\prime \le S_b^\prime$\footnote{See the reference in footnote 2.}, we get
	\begin{equation}
	\label{eq:ineq_DeltaH}
	\Delta \mathcal{H} \leq \Delta S,
	\end{equation}
	which provides an alternative derivation of the inequality $W \le k_BT \Delta S$.
	
	Finally, setting $k_BT=1$ for convenience, we establish the result
	\begin{equation}
	\label{eq:ratio}
	\lim_{\epsilon\rightarrow \delta} \frac{W}{\Delta S} = 1
	\end{equation}
	mentioned near the end of the main text.
	Taking the partial derivatives of the quantities
	\begin{equation}
	W = \Phi \ln \frac{1 + \epsilon}{1-\epsilon}
	\quad \text{and} \quad
	\Delta S = S(\delta^\prime) - S(\delta)
	\end{equation}
	with respect to $\epsilon$, at fixed $\delta$ and $\tau$, we get
	\begin{equation}
	\begin{split}
	\frac{\partial W}{\partial\epsilon} &= \frac{\partial\Phi}{\partial\epsilon} \ln \frac{1 + \epsilon}{1-\epsilon} + \frac{2\Phi}{1-\epsilon^2} \\
	\frac{\partial \Delta S}{\partial\epsilon} &= \frac{\partial\Phi}{\partial\epsilon} \ln \frac{1 + \delta^\prime}{1-\delta^\prime}
	\end{split}
	\end{equation}
	(using $\delta^\prime = \delta - 2\Phi$).
	Along the line $\epsilon=\delta$ we have $W = \Delta S = 0$ as well as
	\begin{equation}
	\frac{\partial W}{\partial\epsilon} = \frac{\partial \Delta S}{\partial\epsilon}.
	\end{equation}
	Eq.~\ref{eq:ratio} then follows by l'H\^{o}pital's rule.

	
	\end{document}